\documentclass[prd,aps,preprint,tightenlines,showpacs,amsmath,amssymb]{revtex4}
\usepackage{graphicx}

\def\mpi2{m_\pi^2}
\def\mK2{m_K^2}

\newcommand{\bea}{\begin{eqnarray}}
\newcommand{\eea}{\end{eqnarray}}
\newcommand{\be}{\begin{equation}}
\newcommand{\ee}{\end{equation}}

\begin{document}
\bibliographystyle{apsrev}

\preprint{CU-TP-1135}


\title{Lattice calculation of SU(3) flavor breaking ratios in $B^{0} - \bar{B}^{0}$ mixing}

\author{Valeriya Gadiyak}
\author{Oleg Loktik}
\email{oleg@phys.columbia.edu}

\affiliation{
Physics Department,
Columbia University,
New York, NY 10027}

\date{\today}

\begin{abstract}
We present an unquenched lattice calculation for the SU(3) flavor breaking ratios of the heavy-light decay constants and the $\Delta B = 2$ matrix elements. The calculation was performed on $16^3 \times 32$ lattices with two dynamical flavors of domain-wall quarks and inverse lattice spacing $1/a = 1.69(5)$ GeV. Heavy quarks were implemented using an improved lattice formulation of the static approximation. In the infinite heavy-quark mass limit we obtain $f_{B_s}/f_{B_d} = 1.29(4)(6)$, $B_{B_s}/B_{B_d} = 1.06(6)(4)$, $\xi = 1.33(8)(8)$ where the first error is statistical and the second systematic.
\end{abstract}

\maketitle

\newpage


\section{Introduction}
\label{sec:intro}
A first-principles, lattice QCD study of $B^{0} - \bar{B}^{0}$ mixing puts the Standard Model to a test and constrains the physics beyond it. In particular, it provides 
a number of theoretical input parameters necessary for constraining the elements of the Cabibbo-Kobayashi-Maskawa (CKM) matrix.
In the Standard Model, the oscillation frequency of $B^{0} - \bar{B}^{0}$ mixing is described by the following expression \cite{Buras:1990fn}:
\begin{equation}\label{eq:ckm}
\Delta m_{q} = \left( \frac{G^{2}_{F}m^{2}_{W}S_{0}}{16\pi^{2}m_{B_{q}}} \right) |V^{*}_{tq}V_{tb}|^{2}\eta_{B}{\mathcal M}_{q},
\end{equation} where $q \in \{d,s\}$, $V_{tq}$ and $V_{tb}$ are CKM matrix elements. Except for the matrix elements and ${\mathcal M}_q$ all quantities in Eq.~(\ref{eq:ckm}) are known from experiments or can be accurately computed by analytic means. The hadronic matrix element
\begin{equation}\label{eq:me}
{\mathcal M}_{q} = \left\langle \bar{B}^{0}_{q}\right\rvert [\bar{b}\gamma^{\mu}(1-\gamma_{5})q] [\bar{b}\gamma_{\mu}(1-\gamma_{5})q]\left\lvert B^{0}_{q} \right\rangle,
\end{equation}
due to its nonperturbative nature is best determined from a lattice calculation.
However, the systematic errors inherent to lattice calculations as well as the uncertainties associated with the perturbative matching between continuum and lattice operators limit the precision with which ${\mathcal M}_{q}$ can be determined. A more promising approach would be to focus not on ${\mathcal M}_{q}$ itself but on the following ratio
\begin{equation}\label{eq:dm_ratio}
\frac{\Delta m_{s}}{\Delta m_{d}} = \left\lvert \frac{V_{ts}}{V_{td}} \right\rvert^{2}\frac{m_{B_{d}}}{m_{B_{s}}}\frac{{\mathcal M}_{s}}{{\mathcal M}_{d}} .
\end{equation}
This ratio relates the matrix element $|V_{td}|$ to the experimentally measurable quantity $\Delta m_s / \Delta m_d$. For historical reasons the hadronic matrix element is often factorized as ${\mathcal M}_{q} = (8/3)m^{2}_{B_q}f^{2}_{B_q}B_{B_q}$. Using this convention we can rewrite Eq.~(\ref{eq:dm_ratio}) as
\begin{equation}
\frac{\Delta m_{s}}{\Delta m_{d}} = \left\lvert \frac{V_{ts}}{V_{td}} \right\rvert^{2}\frac{m_{B_{s}}}{m_{B_{d}}}\xi^{2},
\end{equation}
where
\begin{equation}\label{eq:xi_indirect}
\xi = \frac{f_{B_s}}{f_{B_d}}\sqrt{\frac{B_{B_s}}{B_{B_d}}},
\end{equation}
whose value is directly used in CKM matrix analysis \cite{Ciuchini:2000de}.
Note that since $\xi$ is the ratio of the same quantities with only the mass of the light quark being changed, many systematic errors in the lattice calculation should cancel. Therefore, the focus of this paper will be the determination of the ratios $f_{B_s}/f_{B_d}$, $B_{B_s}/B_{B_d}$ and $\xi$, and not the individual quantities $f_{B}$ and $B_{B}$.
Moreover, we will calculate $\xi$ in two ways. An ``indirect'' approach is to calculate $\xi$ using Eq.~{\ref{eq:xi_indirect}. A ``direct'' approach is to extract
$\xi$ directly from the matrix elements using
\begin{equation}\label{eq:xi_direct}
\xi = \frac{m_{B_d}}{m_{B_s}}\sqrt{\frac{{\mathcal M}_s}{{\mathcal M}_d}}.
\end{equation}

With the notable exceptions \cite{Aoki:2003xb, Gray:2004hd} previous studies of $B^{0} - \bar{B}^{0}$ mixing in lattice QCD have been done in the quenched approximation. It has been estimated \cite{Booth:1994hx,Sharpe:1995qp} that even the values of the ratios, which are less sensitive to systematic errors, are likely to have 10-30\% quenching errors. In order to mitigate the effects of quenching errors, this calculation was done on lattices with two dynamical quarks. 
  
This paper is organized as follows. In Sec.~\ref{sec:action} we discuss the lattice action used in our simulation. In Sec.~\ref{sec:symmetry} we focus on the symmetries of the action. In Sec.~\ref{sec:qcd_matching} we review the connection between continuum QCD and continuum heavy-quark effective theory (HQET). In Sec.~\ref{sec:matching} we discuss the matching between continuum HQET operators and their lattice counterparts. In Sec.~\ref{sec:computation} we review our methods of computing matrix elements on the lattice. We present the results of our calculation in Sec.~\ref{sec:results}.


\section{Action}
\label{sec:action}
It is convenient to separate the action used in our simulation into three different
parts.
\begin{equation}\label{eq:action}
S = S_{heavy} + S_{light} + S_{gauge} ,
\end{equation}
where $S_{heavy}$ and $S_{light}$ describe the heavy- and light-quark sectors and $S_{gauge}$ is the gauge part of the action. In this section we discuss the choices made for each part. Unless it is given explicitly we assume the lattice spacing $a$ to be equal to 1 for the rest of this paper.

                                                                                
\subsection{Heavy Quark Action}
\label{subsec:hqet}
Heavy Quark Effective Theory (HQET) is an expansion of the QCD Lagrangian
in inverse powers of the heavy-quark mass \cite{Eichten:1989zv,Grinstein:1990mj,Georgi:1990um}. 
This expansion is justified as long as the momenta involved are 
small compared to the heavy-quark mass. For B mesons the 
typical momentum transfer is of the order of $\Lambda_{QCD} \approx$ 200MeV.
Since this is in fact much smaller than the mass of the b-quark, we expect that
the leading order of HQET provides a sufficiently accurate description of
B physics phenomena. Thus, we will take the $m_{b} \rightarrow \infty$ limit of HQET,
also known as static-quark approximation. On the lattice the static-quark action is given by
\begin{equation}\label{eq:hqa} S_{heavy} = \sum_{x} \bar{h}(x) \nabla^{-}_{0}h(x), \end{equation} 

\begin{equation}\label{eq:hq_dir} \nabla^{-}_{0}h(x) = h(x) - U_{0}^{\dagger}(x-\hat0)h(x-\hat0), \end{equation}
where $U_{0}(x-\hat0)$ is the gauge link in temporal direction between sites $x$ and $x-\hat0$. The static-quark field $h$ satisfies $\gamma_{0}h = h$. We will discuss the implications and the symmetries associated with this formulation in the next section. Here we just note that the static approximation describes exactly what its name suggests: a heavy quark fixed in space and propagating in the time direction only. The propagator is given explicitly by   
\begin{equation} G_{0}(x,y) = \theta(t_{x} - t_{y})\delta({\mathbf x} - {\mathbf y})\left[U^{\dagger}_{0}({\mathbf y}, t_{y}) U^{\dagger}_{0}({\mathbf y}, t_{y}+1)\ldots  U^{\dagger}_{0}({\mathbf y}, t_{x}-1)\right]. \end{equation}
The lattice static-quark action (\ref{eq:hqa}) is O($a$)-improved as was pointed out in Ref. \cite{Kurth:2000ki}.
 Previous studies \cite{Duncan:1994uq,Ewing:1995ih,Gimenez:1996sk,Christensen:1996sj}, which relied on the action in Eq.~(\ref{eq:hqa}), were plagued by large 
statistical errors inherent to this formulation. However, following the ideas
developed in Ref. \cite{DellaMorte:2003mn}, we use a modified lattice formulation of the static-quark action, which
circumvents the poor signal-to-noise ratio while maintaining the key advantages.
We replace the simple gauge link $U_{0}(x)$ in the backward derivative in Eq.~(\ref{eq:hq_dir}) with an average of the 6 staples around that link
\begin{equation}\label{eq:staple} V_{0}(x) = \frac{1}{6}\sum^{3}_{j=1} \left[ U_{j}(x)U_{0}(x+\hat{j})U^{\dagger}_{j}(x+\hat{0})  + U^{\dagger}_{j}(x-\hat{j})U_{0}(x-\hat{j})U_{j}(x+\hat{0}-\hat{j}) \right]. \end{equation}
 Now our propagator for the heavy quark is given explicitly by 
\begin{equation}\label{eq:hq_imp_prop} G_{h}(x,y) = \theta(t_{x} - t_{y})\delta({\mathbf x} - {\mathbf y})\left[V^{\dagger}_{0}({\mathbf y}, t_{y}) V^{\dagger}_{0}({\mathbf y}, t_{y}+1)\ldots  V^{\dagger}_{0}({\mathbf y}, t_{x}-1)\right]. \end{equation}

\subsection{Light Quark Action}
\label{subsec:dwf}                                                               We will use the domain-wall fermion (DWF) formalism to describe light quarks on the lattice. The DWF formulation achieves improved chiral properties at finite lattice spacing via introduction of an extra dimension \cite{Kaplan:1992bt,Shamir:1993zy}.  
Though the DWF method does not provide exact chiral symmetry, the degree of chiral symmetry breaking and the resulting errors are under control and can be precisely quantified \cite{AliKhan:2000iv,Blum:2000kn}. For a light quark with mass $m_{q}$ the DWF action is
\begin{multline}\label{eq:light_action}
S_{light} = -\sum_{s=0}^{L_{s}-1} a^{4}\sum_{x}\bar\psi_{s}(x)\left\{- \gamma_{\mu}\frac{1}{2}(\nabla^{+}_{\mu} + \nabla^{-}_{\mu}) +\frac{1}{2}\nabla^{-}_{\mu}\nabla^{+}_{\mu} + M_{5} + P_{L}\partial^{+}_{5} - P_{R}\partial^{-}_{5} \right\} \psi_{s}(x) \\ + \sum_{x}m_{q}\bar q(x)q(x).
\end{multline}

Note that the fermion field $\psi_{s}$ has an index $s$, which represents the fifth dimension. The fifth dimension extends from $0$ to $L_{s} -1$ and exact chiral symmetry is achieved in the $L_{s} \rightarrow \infty$ limit. 
Gauge fields are confined to the four-dimensional boundaries at $s=0$ and $L_s-1$ and are present in the forward and backward covariant derivatives  $\nabla^{\pm}_{\mu}$:
\begin{equation}
\nabla^{+}_{\mu}\psi(x) = U_{\mu}(x)\psi(x+\hat{\mu}) - \psi(x) ,
\end{equation}
\begin{equation}
\nabla^{-}_{\mu}\psi(x) = \psi(x) - U^{\dagger}_{\mu}(x-\hat{\mu})\psi(x-\hat{\mu}) . 
\end{equation}
Correspondingly $\partial^{\pm}_{5}$ are derivatives in the fifth dimension with $U_{\mu} = 1$. $P_L = (1-\gamma_{5})/2$ and $P_R = (1+\gamma_{5})/2$ are the projectors for the left- and right-handed spinors. The domain-wall height $M_{5}$ is a parameter of the theory, which we set equal to 1.8. Finally, $q(x)$ represents the physical four-dimensional quark field constructed from the light modes at $s=0$ and $L_s-1$
\begin{equation}
q(x) = P_{L}\psi_{0}(x) + P_{R}\psi_{L_s-1}(x) ,
\end{equation}  
\begin{equation}
\bar{q}(x) = \bar{\psi}_{0}(x)P_R + \bar{\psi}_{L_s-1}(x)P_L .
\end{equation}  
\subsection{Gauge Action}
\label{subsec:gauge} 
The simplest satisfactory formulation of SU(3) gauge fields on the lattice is
given by the standard Wilson plaquette action
\begin{equation}
S^{Wilson}_{gauge} = -\frac{\beta}{3}\sum_{P} {\rm ReTr}[U_{P}], 
\end{equation}
where $\beta=6/g^{2}_{0}$ with $g_{0}$ denoting the bare lattice coupling. $U_{P}$ is the path-ordered product of links around the $1\times1$ plaquette $P$. Recently, it has been realized that it is possible to construct lattice gauge actions that 
are superior to the naive plaquette action in terms of approximating the continuum gauge fields. We will use the doubly-blocked Wilson (DBW2) action \cite{Takaishi:1996xj,deForcrand:1999bi}, which was shown to have excellent chiral properties when combined with the DWF action \cite{Aoki:2002vt}: 
\begin{equation}
S_{gauge} = -\frac{\beta}{3} \left( (1-8c_{1}) \sum_{P} {\rm ReTr}[U_{P}] + c_{1}\sum_{R} {\rm ReTr}[U_{R}] \right) , 
\end{equation}
where $U_{R}$ is the path-ordered product of links around the $1\times2$ rectangle $R$. The parameter $c_{1}$ was set to $-1.4069$ and $\beta = 0.8$ which corresponds to $a^{-1} \approx 1.7$ GeV.

\section{Symmetries of the action}
\label{sec:symmetry}
The advantages of the choices we made for our action in Eq.~(\ref{eq:action}) become clear when we consider its large symmetry group. We focus on two symmetries particularly important for our calculation.

\subsection{Chiral symmetry}
\label{subsec:chi_symm}
${\rm SU}(N_{f})_{L} \otimes {\rm SU}(N_{f})_{R}$ chiral rotations are given by
\begin{equation}\label{eq:chi_rotL}
\psi_{s}(x) \to U_{L}(x)\psi_{s}(x),  \quad  0 \leq s \leq L_{S}/2 -1 ,
\end{equation} 
\begin{equation}\label{eq:chi_rotR}
\psi_{s}(x) \to U_{R}(x)\psi_{s}(x),  \quad  L_{S}/2 \leq s \leq L_{S} -1 ,
\end{equation} 
where $U_{L}(x)$ and $U_{R}(x)$ acting on fermion fields on the left- and right-hand halves of the five-dimensional lattice are given by 
\begin{equation}
U_{L}(x) = e^{i\epsilon^{a}_L(x)T^{a}} , 
\end{equation}
\begin{equation}
U_{R}(x) = e^{-i\epsilon^{a}_R(x)T^{a}} , 
\end{equation}
with $T_{a}$ being the generators of the ${\rm SU}(N_{f})$ flavor group. For finite $L_{s}$ the symmetry due to transformations (\ref{eq:chi_rotL}) and (\ref{eq:chi_rotR}) is not exact. The slight breaking of the chiral symmetry manifests itself through the appearance of an additional mass term, $m_{res}$, in the effective Lagrangian. This residual mass term falls off rapidly with increasing $L_s$ \cite{Blum:2000kn}. For our choice of $L_s = 12$ we observe $m_{res}/m_{s} = 0.03$ or $m_{res} \approx 2.5$ MeV. We will take into account the explicit breaking of chiral symmetry due to $m_{res}$ simply by shifting the values of our light-quark masses from $m_f$ to $m_f + m_{res}$. Note that the additional chiral symmetry breaking due to the appearance of the dimension-five operators is of the order $am_{res}\Lambda_{QCD} \approx 0.5$ MeV. Since it is much smaller than other systematic errors, we will consider the effects of the dimension-five operators to be negligible. 

\subsection{Heavy-quark spin symmetry}
\label{subsec:HQS}
The ${\rm SU}(2)$ heavy-quark spin symmetry (HQS) \cite{Isgur:1989vq,Isgur:1989ed} transformation is defined as 
\begin{equation}
\psi_{h}(x) \to V\psi_{h}(x),
\end{equation}
with
\begin{equation}
V = e^{-i\phi_{i}\epsilon_{ijk}\sigma_{jk}},
\end{equation}
where $\phi_{i}$ is a parameter and $\sigma_{jk} = \frac{i}{2}[\gamma_{i}, \gamma_{j}]$. The heavy-quark spin symmetry simply means that the heavy-light quark interactions are invariant under arbitrary changes of the spin of the heavy quark. 
Note also that due to the heavy-quark field equation $\gamma_0 h = h$, the tensor
density $T^{\mu\nu}$ is not an independent bilinear. In particular, $T^{0j} = V^{j}$ and $T^{ij} = \epsilon^{ijk}A^{k}$.


\section{Matrix elements in the static limit}
\label{sec:qcd_matching}
The expression for the hadronic matrix element in Eq.~(\ref{eq:me}) is a continuum QCD result. It is not obvious how to compute it directly on the lattice, since the inverse lattice spacing $a^{-1} < m_{b}$ for the current generation of computers. In order to make it computable we will make use of HQET. The relevant continuum operator, here normalized according to the $\overline{\rm MS}$(NDR) scheme, is
\begin{equation}\label{eq:msndr_v-a}
O^{\rm \overline{MS}(NDR)}_{(V-A)(V-A)} = \left[\bar{b}\gamma^{\mu}(1-\gamma_{5})q\right]\left[\bar{b}\gamma_{\mu}(1-\gamma_{5})q\right] . 
\end{equation}
This operator is evaluated between parity-even states in Eq.~(\ref{eq:me}). Therefore, only the parity-even part gives a nonzero contribution to the matrix element. The parity-even part 
\begin{equation}\label{eq:msndr_vvaa}
O^{\rm \overline{MS}(NDR)}_{VV+AA} = \left(\bar{b}\gamma^{\mu}q\right)\left(\bar{b}\gamma_{\mu}q\right) + \left(\bar{b}\gamma^{\mu}\gamma_{5}q\right)\left(\bar{b}\gamma_{\mu}\gamma_{5}q\right) , 
\end{equation}
can be written using the operator product expansion as following 
\begin{equation}
O^{\rm \overline{MS}(NDR)}_{VV+AA}(\mu_{b}) = Z_{1}(\mu_{b},\mu) O^{\rm HQET}_{VV+AA}(\mu) +  Z_{2}(\mu_{b},\mu)O^{\rm HQET}_{SS+PP}(\mu) +  {\mathcal O}(1/\mu_{b}). 
\end{equation}
The leading dependence of $O^{\rm \overline{MS}(NDR)}_{VV+AA}$ on the b-quark mass is absorbed into analytic coefficients $Z_{1}(\mu_{b},\mu)$ and $Z_{2}(\mu_{b},\mu)$. Effectively, $Z_{1,2}$ encode the ``short-distance'' physics between scales $\mu_{b} \approx$ 5 GeV and $\mu \approx$ 1.7 GeV whereas $O^{\rm HQET}_{i}(\mu)$ are operators in HQET, which describe the ``long-distance'' physics below $\mu$.   
\begin{equation}\label{eq:hqet_vvaa}
O^{\rm HQET}_{VV+AA} = 2\left(\bar{h}^{(+)}\gamma^{\mu}q\right)\left(\bar{h}^{(-)}\gamma_{\mu}q\right) + 2\left(\bar{h}^{(+)}\gamma^{\mu}\gamma_{5}q\right)\left(\bar{h}^{(-)}\gamma_{\mu}\gamma_{5}q\right),
\end{equation}
\begin{equation}\label{eq:hqet_sspp}
O^{\rm HQET}_{SS+PP} = 2\left(\bar{h}^{(+)}q \right)\left(\bar{h}^{(-)}q\right) + 2\left(\bar{h}^{(+)}\gamma_{5}q\right)\left(\bar{h}^{(-)}\gamma_{5}q\right),
\end{equation}
where
\begin{equation}
h^{(\pm)}(x) = e^{\pm imv\cdot x}\frac{1 \pm \gamma_{\mu}v^{\mu}}{2}b(x) . 
\end{equation}
The full-theory operator in Eq.~(\ref{eq:msndr_vvaa}) can create two heavy quarks or annihilate two heavy antiquarks. Since the field $\bar{h}^{(+)}$ cannot annihilate a heavy antiquark and $\bar{h}^{(-)}$ cannot create a heavy quark we need a compensating factor of $2$ in Eq.~(\ref{eq:hqet_vvaa}) and Eq.~(\ref{eq:hqet_sspp}). Note also that the operators $O^{\rm HQET}_{i}$ having no dependence on the mass of the b-quark are well suited for a static-limit lattice calculation.

We quote the results for $Z_{1,2}$ obtained via perturbative calculations \cite{Gimenez:1992is,Ciuchini:1996sr,Buchalla:1996ys} :  
\begin{multline}
Z_{1}(m_b,\mu) = \left[ \frac{\alpha_{s}(m_b)}{\alpha_{s}(\mu)} \right]^{-\frac{12}{25}} \left(1-14\frac{\alpha_{s}(m_{b})}{4\pi} + \frac{\alpha_{s}(\mu) - \alpha_{s}(m_b)}{4\pi}\tilde{J} \right) \\ +  2\frac{\alpha_{s}(m_b)}{4\pi}\left(\left[ \frac{\alpha_{s}(m_b)}{\alpha_{s}(\mu)} \right]^{-\frac{12}{25}} - \left[ \frac{\alpha_{s}(m_b)}{\alpha_{s}(\mu)} \right]^{-\frac{4}{25}} \right), 
\end{multline}
\begin{equation}\label{eq:z2}
Z_{2}(m_b,\mu) = -8\frac{\alpha_{s}(m_b)}{4\pi}\left[ \frac{\alpha_{s}(m_b)}{\alpha_{s}(\mu)} \right]^{-\frac{4}{25}},
\end{equation} 
where we assumed the number of flavors $n_f$ = 4 for the evaluations of the anomalous dimensions and beta-function coefficients. For $\tilde{J} = \frac{\tilde{\gamma}^{(0)}_{11}\beta_{1}}{2\beta^{2}_{0}} - \frac{\tilde{\gamma}^{(1)}_{11}}{2\beta_{0}}$  we obtain 3.881. According to the latest PDG review \cite{Eidelman:2004wy}  $\Lambda^{(5)}_{QCD}$ = 217 MeV, which implies $\Lambda^{(4)}_{QCD}$ = 276 MeV for our calculation. Then $\alpha_{s}(m_b)$ = 0.213 for $m_b$ = 4.3 GeV and $\alpha_{s}(\mu)$ = 0.309 for $\mu$ = 1.7 GeV. Thus, the numerical values for $Z_{1,2}$ are 
\begin{equation}
Z_{1} = 0.880 , \quad Z_{2} = -0.144 . 
\end{equation}    

Similarly, we can express the pseudoscalar meson decay constant $f_B$ in terms of the HQET matrix element 
\begin{equation}
f_{B}\sqrt{m_{B}} = Z_{0}(m_b, \mu)\langle 0|\bar{h}\gamma_{0}\gamma_{5}q|B\rangle , 
\end{equation}
where $m_B$ is the meson mass and $Z_{0}$ is known to the precision of the two-loop anomalous dimension calculation \cite{Ji:1991pr,Broadhurst:1991fz}. With the number of flavors equal to 4 the result is 
\begin{equation}
Z_{0} = \left[ \frac{\alpha_{s}(m_b)}{\alpha_{s}(\mu)} \right]^{-\frac{6}{25}}
\left(1-\frac{8}{3}\frac{\alpha_{s}(m_{b})}{4\pi} + \frac{\alpha_{s}(\mu) - \alpha_{s}(m_b)}{4\pi}J \right) ,  
\end{equation}
with $J = \frac{\gamma^{(0)}\beta_{1}}{2\beta^{2}_{0}} - \frac{\gamma^{(1)}}{2\beta_{0}}$ = 0.948. Assuming other parameters to be the same as for $Z_{1,2}$ we get
\begin{equation}
Z_{0} = 1.05 . 
\end{equation} 

                                                                                  
\section{Matching continuum and lattice operators}
\label{sec:matching}
In order to relate our lattice results to the physical values of
observables we need to know how to match lattice operators to the effective theory operators defined in Sec.~\ref{sec:qcd_matching}. Such matching depends on the
details of the action used in simulation. The matching coefficients calculated at one-loop level in perturbation theory or nonperturbatively are known for a limited number of actions. Unfortunately, a matching calculation for the combination of improved static
quarks and DWF quarks has not yet been carried out. Therefore, in this paper we will do lattice-to-continuum matching at tree level.

Previous studies involving static heavy quarks and Wilson fermions had significant corrections to the tree-level values of $f_B$ and $B_B$.
Both perturbative \cite{Boucaud:1992nf,Duncan:1994uq} and nonperturbative \cite{Heitger:2003xg,DellaMorte:2005yc} calculations indicate that for our $a^{-1} \approx 1.7$ GeV there is $17-23\%$ downward correction to the tree-level value of $f_B$. For the matrix element and $B_B$ values there are downward $20-37\%$ corrections, depending on the way the perturbative calculation is performed, according to Ref. \cite{Flynn:1990qz,Gimenez:1998mw,Becirevic:2001xt}. Since domain-wall fermions are akin to Wilson fermions, we expect that,
apart from a peculiar wave function renormalization factor discussed below, the corrections to our tree-level values of $f_B$ and $B_B$ will be similar. However, as will be argued below, the uncertainty in matching coefficients has no influence on the precision of $f_{B_s}/f_{B_d}$ and has negligible influence on the values of $B_{B_s}/B_{B_d}$ and $\xi$.
                                                                                  
\subsection{$f_B$ matching}
\label{subsec:fb_match}
For the case of the static heavy-quark and Wilson light-quark action the tree-level
matching for the decay constant would be trivial. The matching coefficient $Z_q$ in Eq.~(\ref{eq:fb_match}) would be 1. In the case of domain-wall quarks,
the physical modes extend slightly into the fifth dimension; therefore,
the overlap between the interpolating fields on the boundary and the physical
modes is different from 1 even at tree level \cite{Aoki:1998vv}. Fortunately, the quark field renormalization factor, which includes this overlap effect,
was calculated nonperturbatively in Ref \cite{Blum:2001sr}. With $q$ representing the renormalized light-quark field we have
\begin{equation}
q = Z_q^{1/2} q^{\rm latt}.
\end{equation}
For our value of $M_5$ = 1.8 the $Z_q$ determined from the conserved axial current is
\begin{equation}
Z_q^{1/2} = 0.897(11) .
\end{equation}
Combining this result with the QCD-HQET matching described in Sec.~\ref{sec:qcd_matching} we obtain the final formula for $f_B$
\begin{equation}\label{eq:fb_final}
f_B\sqrt{m_B} = Z_0 Z_q^{1/2} \Phi^{\rm latt}_B a^{-3/2} ,
\end{equation}
where $\Phi^{\rm latt}_B$ is the lattice counterpart of the continuum quantity
$f_B\sqrt{m_B}$. The numerical value of $\Phi^{\rm latt}_B$ can be obtained via Eq.~(\ref{eq:f_b_latt}). Note that by including $Z_q$ in Eq.~(\ref{eq:fb_final}) we deviate from a strictly tree-level calculation. However, we stress that the $Z_q$ factor is specific to the domain-wall fermion formulation and contains a large tree-level piece. By including it we compensate for the fifth dimension and make our results more easily comparable to the studies done with Wilson fermions. Except for $Z_q$ there are no other renormalization factors, since all
four-dimensional amplitudes are evaluated at tree level.
\subsection{Four-fermion operator matching}
\label{subsec:4f_op_match}
The tree-level matching of four-fermion operators is straightforward.
There is no operator mixing and we only have to take into account
the light-quark overlap factor. Thus
\begin{equation}
O^{\rm HQET}_{i}(a^{-1}) = Z_q O^{\rm latt}_{i} .
\end{equation}
Combining this with the QCD-HQET matching discussed in Sec.~\ref{sec:qcd_matching}
we obtain
\begin{equation}\label{eq:full_match}
O^{\rm \overline{MS}(NDR)}_{VV+AA} =  Z_1 Z_q O^{\rm latt}_{VV+AA} + Z_2 Z_q O^{\rm latt}_{SS+PP} .
\end{equation}
Note that $Z_q$ cancels in the $B_B$ determination since
\begin{equation}\label{eq:b_b_mix}
B_B = \frac{\langle \bar{B} | O^{\rm \overline{MS}(NDR)}_{VV+AA} | B \rangle}{\frac{8}{3} f^{2}_B m^2_B} = Z_1Z^{-2}_0 B^{\rm latt}_{VV+AA} + Z_2Z^{-2}_0 B^{\rm latt}_{SS+PP} ,
\end{equation}
where
\begin{equation}\label{eq:b_latt_def}  
B^{\rm latt}_{O_i} \equiv \frac{\langle \bar{B}|O^{\rm latt}_{i}(0)|B\rangle }{\frac{8}{3}(\Phi^{\rm latt}_B)^2m_B} . 
\end{equation} 
Even though we match at tree level it is worthwhile to consider
the effect of a one-loop matching. First, we observe that since $Z_{2}$ in
Eq.~(\ref{eq:z2}) is already order $\alpha_s$, the lattice matching at 
tree level is sufficient for $O^{\rm HQET}_{SS+PP}$.
The operator $O^{\rm HQET}_{VV+AA}$ needs to be matched at one-loop level. Because of chiral symmetry, $O_{VV+AA}$ can mix only with $O_{SS+PP}$ and $O_{TT}$. As discussed in Sec.~\ref{sec:symmetry}, $O_{TT}$ can be decomposed into $O_{SS+PP}$ and $O_{VV+AA}$. Therefore, $O_{VV+AA}$ can mix with $O_{SS+PP}$ only. The perturbative calculations for $O_{VV+AA}$ with Wilson quarks \cite{Flynn:1990qz,Gimenez:1998mw} show that it does not mix with $O_{SS+PP}$. Thus, $O_{VV+AA}$ should be renormalized by a single multiplicative constant. For $a^{-1} \approx 2$ GeV the typical value of this constant is 0.59. This factor would multiply $Z_1Z_q$ in Eq.~(\ref{eq:full_match}) and $Z_1$ in Eq.~(\ref{eq:b_b_mix}). With this correction our tree-level results for ${\mathcal M}$ and $B_B$ would decrease by 37\%, but the results for the ratios ${\mathcal M}_s/{\mathcal M}_d$ and $B_{B_s}/B_{B_d}$ would change by less than 1\%. Thus, for the purposes of the ratios determination the tree-level matching is sufficiently accurate. 


\section{Computation of matrix elements on the lattice}
\label{sec:computation}
Our method for computing the relevant matrix elements on the lattice is fairly 
standard. We calculate the following two- and three-point correlation functions 
\begin{equation}\label{eq:two_pb}
{\mathcal C}^{PC}(t, t_0) = \sum_{\mathbf x}\langle0|A^{P}_{0}({\mathbf x},t)A^{C}_{0}(t_0)^{\dagger}|0\rangle , 
\end{equation}
\begin{equation}\label{eq:two_bb}
{\mathcal C}^{CC}(t, t_0) = \langle0|A^{C}_{0}(t)A^{C}_{0}(t_0)^{\dagger}|0\rangle , 
\end{equation}
\begin{equation}\label{eq:three_bb}
{\mathcal C}_{O_{i}}(t_{1},t) = \sum_{\mathbf x}\langle0|\bar{A}^{C}_{0}(t_{1})O^{\rm latt}_{i}({\mathbf x},t)A^{C}_{0}(0)^{\dagger}|0\rangle , 
\end{equation}
with
\begin{equation}
A^{P}_{0}({\mathbf x}, t) = \bar{h}({\mathbf x}, t)\gamma_{0}\gamma_{5}q({\mathbf x}, t) , 
\end{equation}
\begin{equation}
A^{C}_{0}(t) = \sum_{{\mathbf(x,y)} \in C}\bar{h}({\mathbf x},t)\gamma_{0}\gamma_{5}q({\mathbf y},t) , 
\end{equation}
\begin{equation}
\bar{A}^{C}_{0}(t) = \sum_{{\mathbf(x,y)} \in C}\bar{q}({\mathbf x},t)\gamma_{0}\gamma_{5}h({\mathbf y},t) , 
\end{equation}
where C and P stand for cube-smeared and point interpolation operators used for B mesons. The smearing was done in a cube with side $L_b$ = 9, since it showed the best isolation of the B-meson ground state. The sources were located at $t_0 = 0$ and $t_1 = 20$. We used Coulomb gauge fixing for cube-smeared sources and sinks.

The heavy-light two-point correlators for large t behave as 
\begin{equation}
{\mathcal C}^{PC}(t, 0) \xrightarrow{ t \gg 0} {\mathcal A}^{PC} e^{-Et} , 
\end{equation}
\begin{equation}
{\mathcal C}^{CC}(t, 0) \xrightarrow{ t \gg 0} {\mathcal A}^{CC} e^{-Et} . 
\end{equation}
We obtain ${\mathcal A}^{PC}$ and ${\mathcal A}^{CC}$ by simultaneous fit of ${\mathcal C}^{PC}(t)$ and ${\mathcal C}^{CC}(t)$ to an exponential function with $E$ being the unphysical binding energy for the lightest pseudoscalar state. The quality of the ground state isolation can be monitored via plots of $E(t)$ and ${\mathcal C}_{O_i}(20,t)$. As can be seen
in FIG.~\ref{fig:gnd_state} we have good ground state isolation starting at $t \geqslant 6$. Once  ${\mathcal A}^{PC}$ and ${\mathcal A}^{CC}$ are determined we can calculate the matrix element of the lattice static-light axial current  
\begin{equation}\label{eq:f_b_latt}
\Phi^{\rm latt}_B = \langle0|A^{P}_{0}(0)|B\rangle = \frac{\sqrt{2}{\mathcal A}^{PC}}{\sqrt{{\mathcal A}^{CC}}} .
\end{equation}
The continuum HQET matrix element is then directly proportional to its lattice value
\begin{equation}\label{eq:fb_match}
\langle 0|\bar{h}\gamma_{0}\gamma_{5}q|B\rangle = Z^{1/2}_q \Phi^{\rm latt}_B ,
\end{equation} 
where $Z_q$ is a matching factor, which was discussed in Sec.~\ref{sec:matching}.

Similarly, the three-point function with sources for both heavy and light quarks located at $0$ and $t_{1}$ behaves as 
\begin{equation}
{\mathcal C}_{O_{i}}(t_{1},t) \xrightarrow{t_1 \gg t \gg 0} \frac{1}{2m_B} {\mathcal A}^{CC} 
\langle \bar{B}|O^{\rm latt}_{i}(0)|B\rangle e^{-Et_{1}} .
\end{equation}
The most elegant way to extract the matrix element would be to divide ${\mathcal C}_{O_i}(t_1,t)$ by ${\mathcal C}^{CC}(t_1)$. Unfortunately, for $t_1 = 20$ the smeared-sink two-point correlation function exhibits prohibitively large errors. We will circumvent this difficulty by using
two two-point correlation functions at earlier times    
\begin{equation}\label{eq:me_def}
M_{O_i}(t_1,t) \equiv \frac{2{\mathcal A}^{CC}{\mathcal C}_{O_{i}}(t_{1},t)}{{\mathcal C}^{CC}(t,t_{1}){\mathcal C}^{CC}(t,0)} \xrightarrow{t_1 \gg t \gg 0} \frac{1}{m_B} \langle \bar{B}|O^{\rm latt}_{i}(0)|B\rangle .  
\end{equation}
The raw data for $M_{VV+AA}$ and $M_{SS+PP}$ are depicted in FIG.~\ref{fig:me_vvaa} and FIG.~\ref{fig:me_sspp}. 
Finally, the lattice B parameter is calculated as follows
\begin{equation}\label{eq:b_b_def}
R_{O_i}(t_1,t) \equiv \frac{{\mathcal C}_{O_i}(t_1,t)}{\frac{8}{3}{\mathcal C}^{PC}(t,t_1){\mathcal C}^{PC}(t,0)}  \xrightarrow{t_1 \gg t \gg 0} B^{\rm latt}_{O_i} . 
\end{equation}
As can be seen in FIG.~\ref{fig:b_vvaa} and FIG.~\ref{fig:b_sspp} the errors for $R_{O_i}$ are smaller than for 
$M_{O_i}$, since the point-sink two-point correlation functions are less noisy. 


\section{Numerical results}
\label{sec:results}
We performed measurements on $16^{3} \times 32$ lattices with two dynamical flavors of domain-wall quarks. 
The inverse lattice spacing determined from the $\rho$-meson mass corresponds to
1.691(53) GeV, which is consistent with the value of 1.688(21) GeV obtained from
the Sommer scale.
In order to make the chiral extrapolation all quantities
were calculated at three values of the dynamical $u$ and $d$ quarks with the bare masses:
0.02, 0.03 and 0.04. The bare quark mass 0.0446(29) corresponds to the mass of the strange quark $m_{s}$. The value for $m_{res}$ incorporating the slight breaking of the chiral symmetry was found to be 0.03$m_{s}$. 
The sea and valence quark masses were set to be identical. The pseudoscalar pion mass for the three quark masses is 0.2910(24), 0.3568(25) and 0.4086(21) in lattice units. We performed calculations on 449, 425 and 488 configurations for $m_f = 0.02$, $m_f = 0.03$ and $m_f = 0.04$ respectively. Configurations were chosen at random from 5,000 thermalized hybrid Monte Carlo trajectories separated by 10 trajectories. Statistical errors were computed using the jackknife method with block size equal to 5 configurations. Static-quark propagators were computed according to Eq.~(\ref{eq:staple}) and Eq.~(\ref{eq:hq_imp_prop}). For the light quarks, all Dirac matrix inversions were performed using the conjugate gradient algorithm with a stopping condition of $10^{-8}$. We used periodic boundary conditions in the spatial directions and antiperiodic in the time direction.
More information on $m_{res}$, conjugate gradient algorithm, $a^{-1}$ determination and lattices generation can be found in Ref. \cite{Aoki:2004ht}.

Now we quote our results.  
The numerical values for the lattice $\Phi^{\rm latt}_B$ matrix element defined in
 Eq.~(\ref{eq:f_b_latt}) are given in Tab.~\ref{tab:f_b} for each value of the bare quark mass. The linear extrapolation to the chiral limit gives $\Phi^{\rm latt}_{B_d}=0.274(8)(22)$. Here and in all following results, the number in first parenthesis always represents statistical error, the numbers in other parentheses represent systematic errors which will be discussed in detail at the end of this section. The value at the physical strange quark mass is $\Phi^{\rm latt}_{B_s}=0.356(8)(29)(5)$. Matching the lattice and continuum operators using Eq.~(\ref{eq:fb_final}) we obtain $f_{B_d}\sqrt{m_{B_d}} = 0.568(17)(50)(^{+0}_{-131})$GeV$^{3/2}$, which corresponds to $f_{B_d} = 247(7)(22)(^{+0}_{-57})$ MeV where we used $m_{B_d}=5279$ MeV (see Ref. \cite{Eidelman:2004wy}). We plotted $f_B\sqrt{m_B}$ in physical units as a function of mass in FIG.~\ref{fig:f_b}.

We quoted the $f_{B_d}$ value solely for the purpose of cross-checking our results with previous calculations in the static limit. Duncan et al. \cite{Duncan:1994uq}  report $f_{B_d} = 188(23)(^{+55}_{-29})$ MeV with the lattice-to-continuum matching factor $\tilde{Z} = 0.77$ at $a^{-1} = 1.78(9)$ GeV. We compensate for $\tilde{Z}$ to make a comparison at tree level and obtain the central value of $f_{B_d} = 244$ MeV, which is quite similar to our result.

Our main focus is the determination of the SU(3) breaking ratios which we can determine much more precisely. For the ratio of the decay constants we obtain 
\begin{equation}
\frac{f_{B_s}}{f_{B_d}} = \frac{\Phi^{\rm latt}_{B_s}}{\Phi^{\rm latt}_{B_d}}\sqrt{\frac{m_{B_d}}{m_{B_s}}} = 1.29(4)(4)(2) ,
\end{equation}
where we assumed $m_{B_s}/m_{B_d} = 1.017$ (see Ref. \cite{Eidelman:2004wy}). 
Our value is somewhat larger than the conventional value $f_{B_s}/f_{B_d} = 1.15$. Interestingly, McNeile and Michael \cite{McNeile:2004wn} also using an
improved lattice formulation of static approximation and unquenched clover action obtained
$f_{B_s}/f_{B_d} = 1.38(13)(8)$.  

The numerical values of $M_{O_i}$ defined in Eq.~(\ref{eq:me_def}) are given in Tab.~\ref{tab:m_oi} in lattice units. The physical hadronic matrix element ${\mathcal M}_q$ can be obtained from $M_{O_i}$ via Eq.~(\ref{eq:full_match}). We show the dependence of ${\mathcal M}_q m^{-1}_{B_{q}}$ on the mass of the light quark in FIG.~\ref{fig:me}. 
The linear extrapolation to the chiral limit gives ${\mathcal M}_d m^{-1}_{B_d} = 0.727(96)(65)(^{+0}_{-269})$ GeV$^3$ and to the mass of the strange quark ${\mathcal M}_s m^{-1}_{B_s} = 1.73(9)(15)(^{+0}_{-64})(7)$ GeV$^3$. 
Thus, for the ratio of matrix elements we obtain
\begin{equation}
\frac{{\mathcal M}_s}{{\mathcal M}_d} = 2.42(32)(21)(10) .
\end{equation} 
To our knowledge, this is the first unquenched calculation that obtained the ratio of matrix elements directly. To put our result in context we quote results
from a previous quenched study.  Bernard, Blum and Soni \cite{Bernard:1998dg}, using extrapolated Wilson action found ${\mathcal M}_s/{\mathcal M}_d = 1.76(10)$ and $2.21(42)$ for constant and linear continuum extrapolations, respectively.
Using Eq.~(\ref{eq:xi_direct}) we calculate $\xi$ directly from the ratio of the matrix elements 
\begin{equation}
\xi = 1.53(10)(7)(3) .
\end{equation}
Finally, the values for $B^{\rm latt}_{O_i}$ obtained using Eq.~(\ref{eq:b_b_def}) can be found in Tab.~\ref{tab:r_oi}. The dependence of $B_B$ obtained using Eq.~(\ref{eq:b_b_mix}) on the mass is depicted in FIG.~\ref{fig:b_b}. Again using the linear extrapolation we obtain $B_{B_d} = 0.812(48)(67)(^{+0}_{-300})$ and $B_{B_s} = 0.864(28)(71)(^{+0}_{-320})(3)$. Thus 
\begin{equation}
\frac{B_{B_s}}{B_{B_d}} = 1.06(6)(3)(1) .
\end{equation}
Combining the above ratio with the ratio of the decay constants we calculate $\xi$ in an ``indirect'' way via Eq.~(\ref{eq:xi_indirect}) 
\begin{equation}\label{eq:xi_num_ind}
\xi = 1.33(8)(6)(2) .
\end{equation}
As can be seen in FIG.~\ref{fig:f_b} and FIG.~\ref{fig:b_b} the separate linear extrapolations for $f_B \sqrt{m_B}$ and $B_B$ fit data points better than the linear fit for ${\mathcal M} m^{-1}_B$ in FIG.~\ref{fig:me}. Therefore, the ``indirect'' method seems to be more reliable than the ``direct'' method for the case of linear chiral extrapolation. Thus, we adopt as our final answer for $\xi$ the value in Eq.~(\ref{eq:xi_num_ind}). 

Now let us discuss errors. The first error is statistical and is easily
quantified using the jackknife method. The number in second parenthesis is the finite volume error and the finite lattice spacing error added in quadrature. As argued in Ref. \cite{Arndt:2004bg} finite volume effects for the above quantities are small ($\leqslant 2\%$) as long as the calculation is 
unquenched. Finite volume errors are being estimated using $\chi$PT which we believe should be accurate for mesons in our volume. The finite lattice spacing errors present a larger concern. Though the action we use is $O(a)$-improved, the operators themselves are not. Therefore, $O(a)$ errors are possible for the individual quantities.  A continuum extrapolation reported in Ref. \cite{Lellouch:2000tw} suggests that the error can be as large as $8\%$. Note that these estimates of finite lattice spacing errors themselves contain quenching errors which could be large. For the individual quantities the biggest uncertainty by far comes from the corrections to the tree-level
matching between the continuum and lattice operators. As discussed in Sec.~\ref{sec:matching} we expect these corrections to be similar to the ones in the case of static heavy quarks and Wilson fermions. We adopt the largest published corrections \cite{Duncan:1994uq,Flynn:1990qz} to tree-level as our error estimates. Thus, the uncertainty associated with the lattice-to-continuum matching is $23\%$ for $f_B$ and $37\%$ for the matrix elements and $B_B$. This matching error is quoted in the third parenthesis for $f_B$, ${\mathcal M}$ and $B_{B}$. For the quantities involving the strange quark, there is an uncertainty associated with the value of the strange quark mass. It is quoted in the third parenthesis for the ratios. It is quoted in the fourth parenthesis for $f_{B_s}$, ${\mathcal M}_{s}$ and $B_{B_s}$.

Finally, let us discuss the uncertainty associated with chiral extrapolation. The chiral correction to $f_B$ predicted via $\chi$PT is well known \cite{Grinstein:1992qt,Sharpe:1995qp}. Here, for the case of $N_f=2$ we have
\begin{equation}\label{eq:f_b_chipt}
f_B = f_0 \left[ 1 - \frac{3(1+3g^2)}{4}\frac{m^2_{PS}}{(4\pi f)^2} \ln\frac{m^2_{PS}}{\mu^2} + \cdots \right] , 
\end{equation}
where $g$ is the heavy-meson coupling to pions. The same expression with $-(1-3g^2)/2$ instead of
$3(1+3g^2)/4$ is predicted for $B_B$. A possible approach to chiral extrapolation is to fit
the lattice data to Eq.~(\ref{eq:f_b_chipt}) and then to extrapolate to small values of pion mass. However, as was pointed out in Ref. \cite{Sanz-Cillero:2003fq} such an approach is misleading since Eq.~(\ref{eq:f_b_chipt}) is not valid at large values of pion mass ($m_{PS} > 500$ MeV). On the other hand, we realize that the linear chiral extrapolation used in our calculation is not satisfactory either, since it ignores legitimate corrections coming from long-distance physics \cite{Kronfeld:2002ab,Becirevic:2002mh}. Clearly, the issue can be resolved only when more data points at lower values of quark masses become available. In our case, an attempt to quantify the chiral extrapolation error via $\chi$PT would border on speculation, since our data is consistent with the linear extrapolation and it is outside the region of validity of $\chi$PT. Therefore, we do not include the uncertainty associated with the chiral extrapolation in our estimation of systematic error. 

Our calculation was performed in the limit of the infinite heavy-quark mass. In principle, the finiteness of the heavy-quark mass can be taken into account by including $1/m_B$ corrections. The naive expectation is that the corrections to the individual quantities should be of the order $\Lambda_{QCD}/m_B$. Then the corrections to the values of the ratios should not exceed $\Lambda_{QCD}/m_B$ times the difference between the value of the ratio and $1$. For example, in the case of $\xi$, the corrections can be of the order $(\xi -1)\Lambda_{QCD}/m_B$. Thus, if $1/m_B$ corrections are taken into account, the change in the value of $B_B$ ratio should not be noticeable, whereas the values of the $f_B$ ratio and $\xi$ could be modified by $2\%-4\%$.

In summary, our final answers for SU(3) flavor breaking ratios are: $f_{B_s}/f_{B_d} = 1.29(4)(6)$, $B_{B_s}/B_{B_d} = 1.06(6)(4)$, $\xi = 1.33(8)(8)$ where the number in the first parenthesis is statistical error and the number in the second parenthesis is the arithmetic sum of systematic errors. Note that these systematic errors do not include the uncertainty associated with chiral extrapolation and possible $1/m_B$ corrections. 
\section{Conclusions}
\label{sec:conclusions}
In this paper we have done an unquenched two-flavor calculation of SU(3) flavor breaking ratios in $B^{0}-\bar{B}^{0}$ mixing. A chirally improved, domain-wall fermion action was used for the light quarks. An improved lattice formulation of the static action was used for the heavy quarks. Two flavors of dynamical quarks were simulated at three values of quark mass with the lowest value equal to half of the physical strange quark mass. Lattice-to-continuum matching of operators was done at tree level, which resulted in large uncertainties for individual quantities, but had negligible influence on the value of the ratios due to the large symmetry groups of the action. Our results suggest that the conventional values for phenomenologically relevant parameters $f_{B_s}/f_{B_d}$ and $\xi$ may be too low.
                                                                                  
\section*{Acknowledgments}
We would like to thank N. Christ, A. Soni, C. Dawson and S. Cohen for useful discussions. The calculation was carried out using RBC Collaboration lattices. This work was supported by DOE grant DE-FG02-92ER40699 and we thank RIKEN, BNL and the U.S. DOE for providing the facilities essential for the completion of this work.
                                                                                  

\bibliography{paper}



\begin{table}[htbp]
\caption{\label{tab:f_b}The values of $\Phi^{\rm latt}_B$ defined in Eq.~(\ref{eq:f_b_latt}) for different light quark masses. Statistical errors are given in parentheses.}
\begin{ruledtabular}
\begin{center}
\begin{tabular}{c c}
$m_f + m_{res}$ & $\Phi^{\rm latt}_B$ \\
\hline                  
0.0214 & 0.314(4) \\
0.0314 & 0.330(4) \\
0.0414 & 0.351(4) \\

\end{tabular}
\end{center}
\end{ruledtabular}
\end{table}    

\begin{table}[htbp]
\caption{\label{tab:m_oi}The values of lattice matrix elements $M_{O_i}$ defined in Eq.~(\ref{eq:me_def}) for different light quark masses. Statistical errors are given in parentheses.}
\begin{ruledtabular}
\begin{center}
\begin{tabular}{c c c}
$m_f + m_{res}$ & $M_{VV+AA}$ & $M_{SS+PP}$ \\
\hline                  
0.0214 & 0.326(13) & -0.206(7) \\
0.0314 & 0.366(11) & -0.235(7) \\
0.0414 & 0.446(12) & -0.276(7) \\

\end{tabular}
\end{center}
\end{ruledtabular}
\end{table}    

\begin{table}[htbp]
\caption{\label{tab:r_oi}The values of $B^{\rm latt}_{O_i}$ defined in Eq.~(\ref{eq:b_latt_def}) and obtained via Eq.~(\ref{eq:b_b_def}) for different light quark masses. Statistical errors are given in parentheses.}
\begin{ruledtabular}
\begin{center}
\begin{tabular}{c c c}
$m_f + m_{res}$ & $B^{\rm latt}_{VV+AA}$ & $B^{\rm latt}_{SS+PP}$ \\
\hline                  
0.0214 & 0.960(28) & -0.605(15) \\
0.0314 & 0.950(18) & -0.613(10) \\
0.0414 & 0.987(16) & -0.611(9) \\

\end{tabular}
\end{center}
\end{ruledtabular}
\end{table}


%
%

\begin{figure}
\includegraphics[width=6.4in]{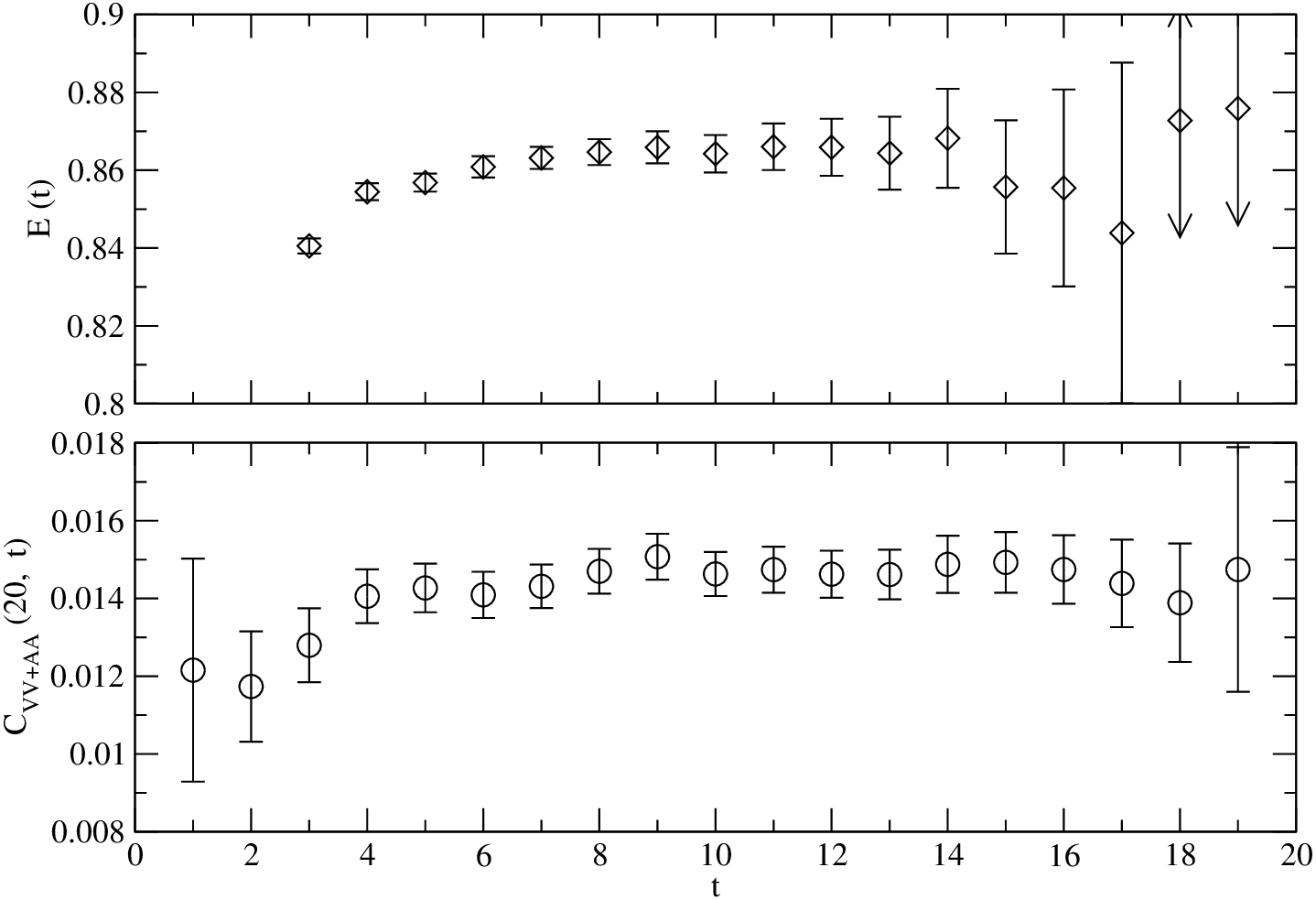}
\caption{\label{fig:gnd_state}The binding energy for the lightest pseudoscalar state $E(t) = ln\frac{C^{LS}(t)}{C^{LS}(t+1)}$ and the three-point correlation function $C_{VV+AA}(20, t)$ as functions of time. Both plots are for the bare light quark mass $m_f = 0.04$. Error bars represent statistical errors.}
\end{figure}

\begin{figure}
\includegraphics[width=6.4in]{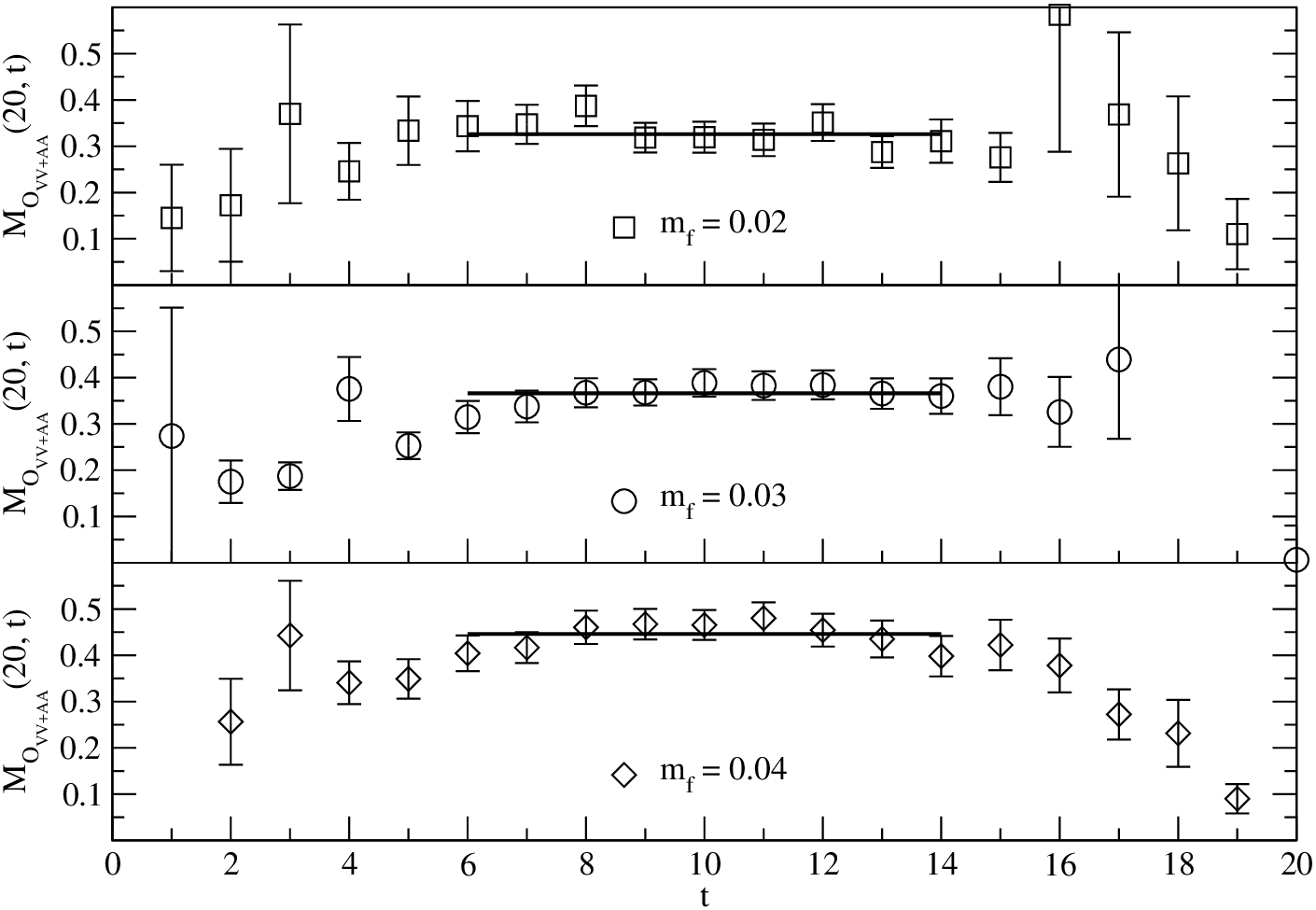}
\caption{\label{fig:me_vvaa}A plot showing the raw data for $M_{O_{VV+AA}}(20,t)$ defined in Eq.~(\ref{eq:me_def}) and Eq.~(\ref{eq:hqet_vvaa}) for different values of light quark masses. Error bars represent statistical errors. Plateaus indicate the weighted averages for $6 \leqslant t \leqslant 14$. Their values are quoted in Tab.~\ref{tab:m_oi}.}
\end{figure}

\begin{figure}
\includegraphics[width=6.4in]{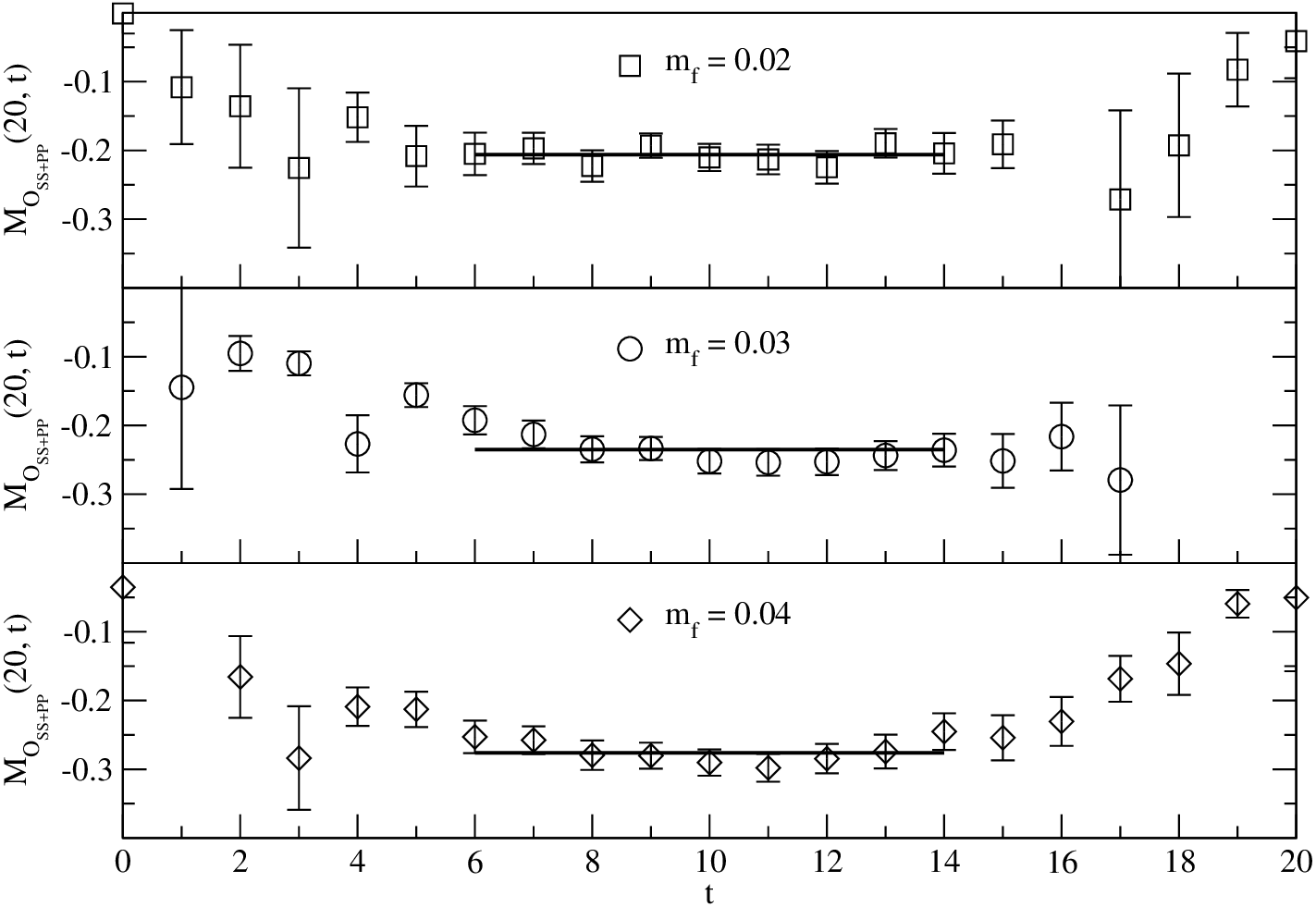}
\caption{\label{fig:me_sspp}A plot showing the raw data for $M_{O_{SS+PP}}(20,t)$ defined in Eq.~(\ref{eq:me_def}) and Eq.~(\ref{eq:hqet_sspp}) for different values of light quark masses. Error bars represent statistical errors. Plateaus indicate the weighted averages for $6 \leqslant t \leqslant 14$. Their values are quoted in Tab.~\ref{tab:m_oi}.}
\end{figure}

\begin{figure}
\includegraphics[width=6.4in]{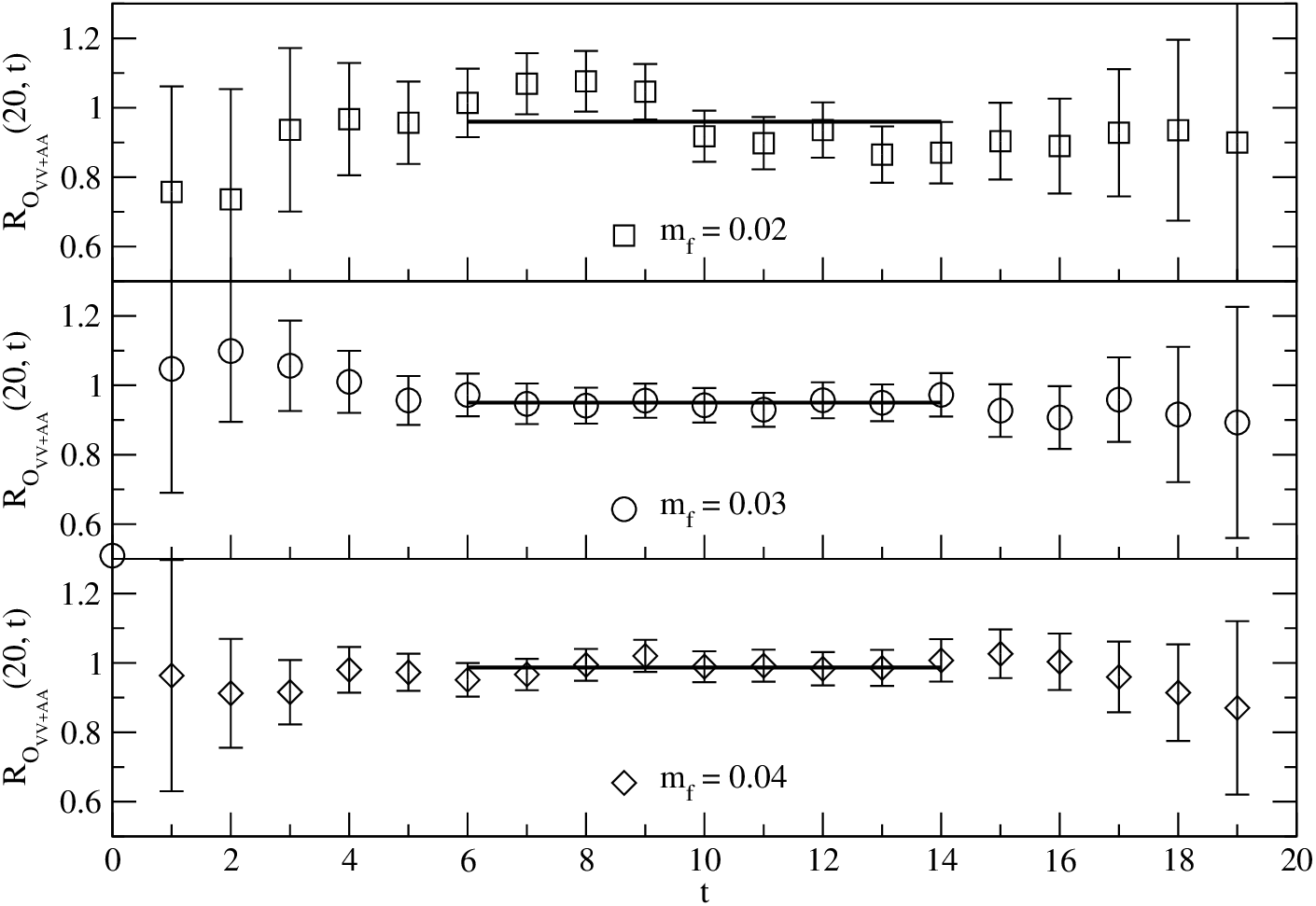}
\caption{\label{fig:b_vvaa}A plot showing the raw data for $R_{O_{VV+AA}}(20,t)$ defined in Eq.~(\ref{eq:b_b_def}) and Eq.~(\ref{eq:hqet_vvaa}) for different values of light quark masses. Error bars represent statistical errors. Plateaus indicate the weighted averages for $6 \leqslant t \leqslant 14$. Their values are quoted in Tab.~\ref{tab:r_oi}.}
\end{figure}

\begin{figure}
\includegraphics[width=6.4in]{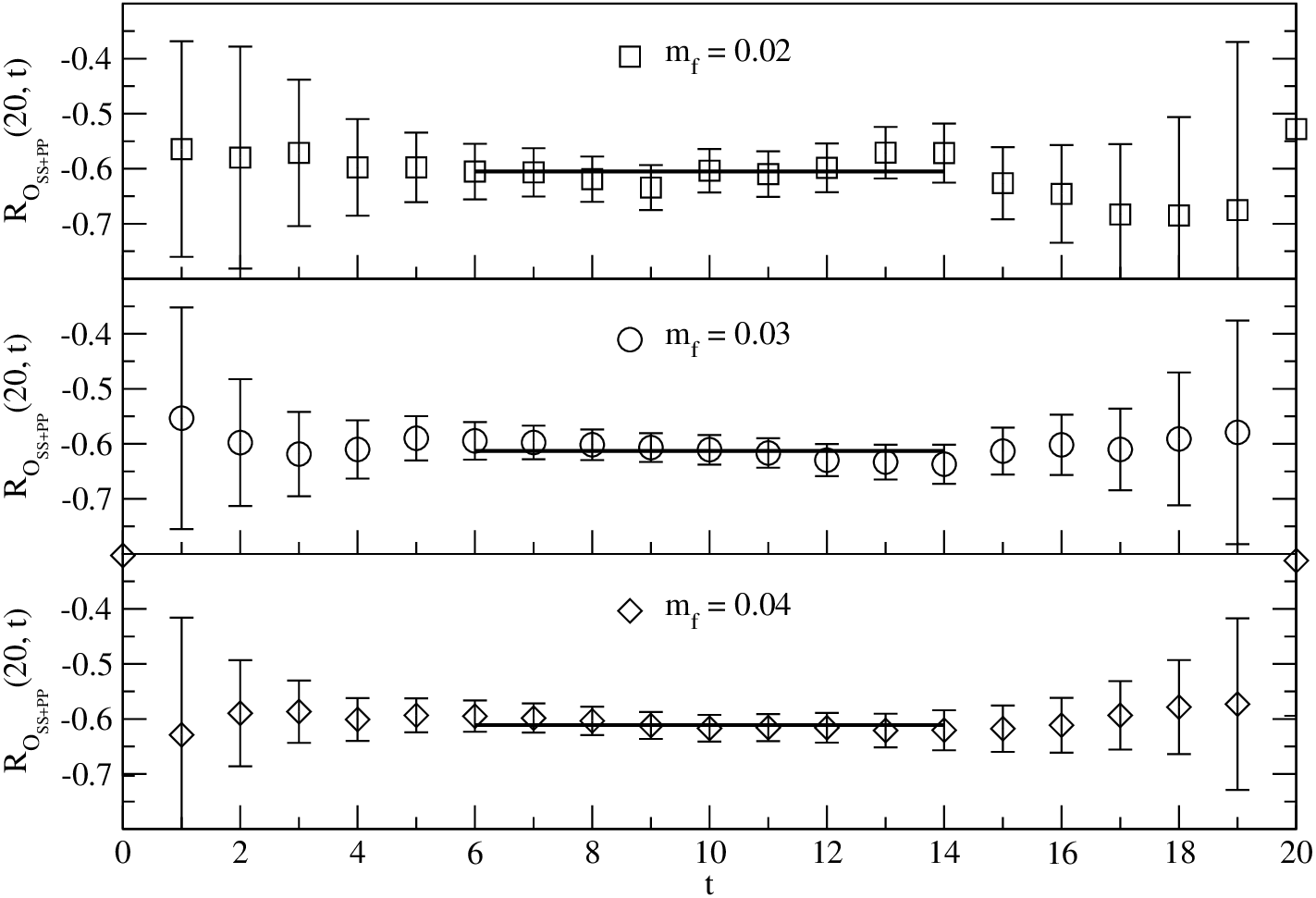}
\caption{\label{fig:b_sspp}A plot showing the raw data for $R_{O_{SS+PP}}(20,t)$ defined in Eq.~(\ref{eq:b_b_def}) and Eq.~(\ref{eq:hqet_sspp}) for different values of light quark masses. Error bars represent statistical errors. Plateaus indicate the weighted averages for $6 \leqslant t \leqslant 14$. Their values are quoted in Tab.~\ref{tab:r_oi}.}
\end{figure}

\begin{figure}
\includegraphics[width=6.4in]{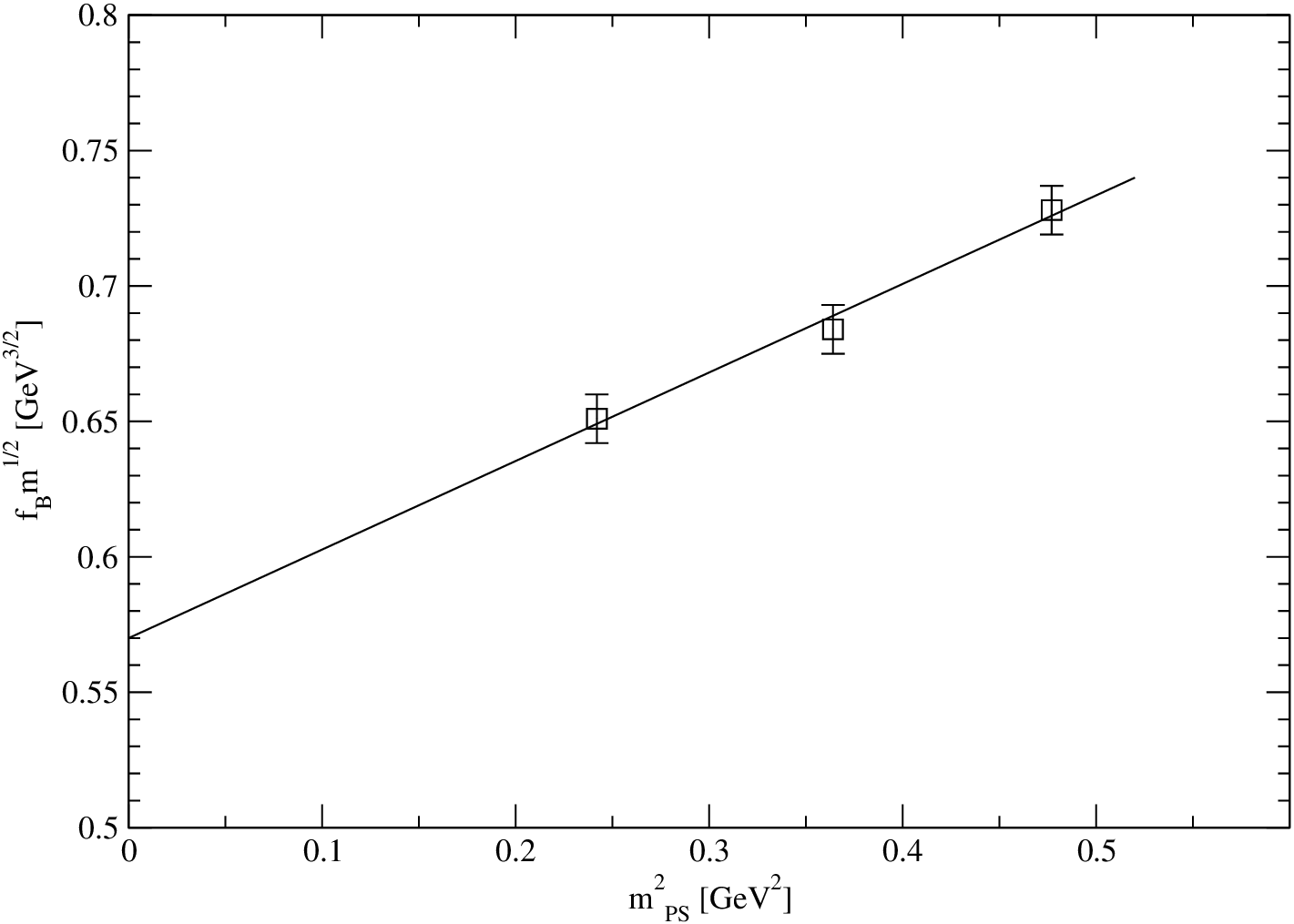}
\caption{\label{fig:f_b}Chiral extrapolation of $f_B\sqrt{m_B}$ obtained using Eq.~(\ref{eq:fb_final}). Error bars represent statistical errors.}
\end{figure}

\begin{figure}
\includegraphics[width=6.4in]{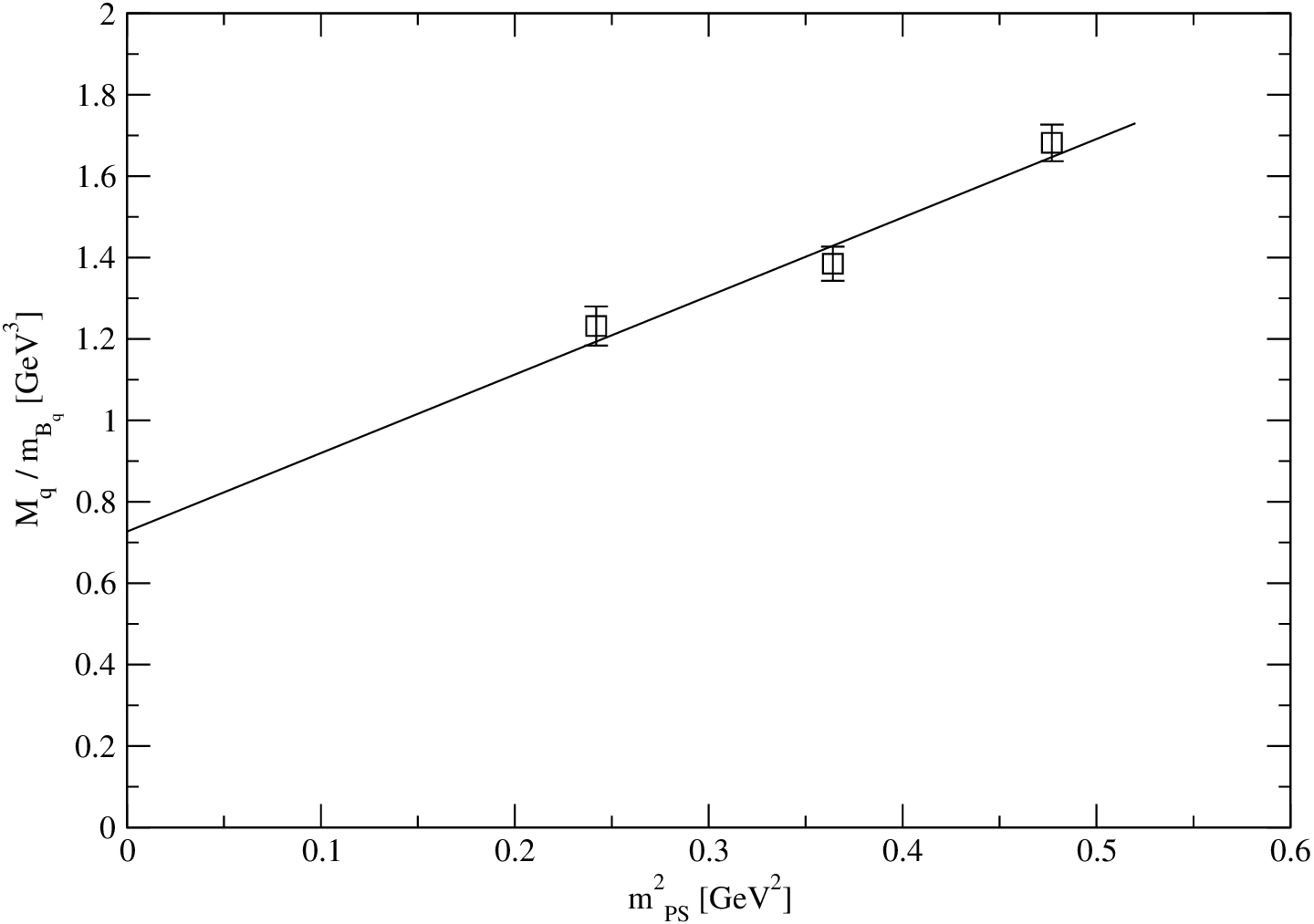}
\caption{\label{fig:me}Chiral extrapolation of ${\mathcal M}_q m^{-1}_{B_q}$ obtained using Eq.~(\ref{eq:full_match}). Error bars represent statistical errors.}
\end{figure}

\begin{figure}
\includegraphics[width=6.4in]{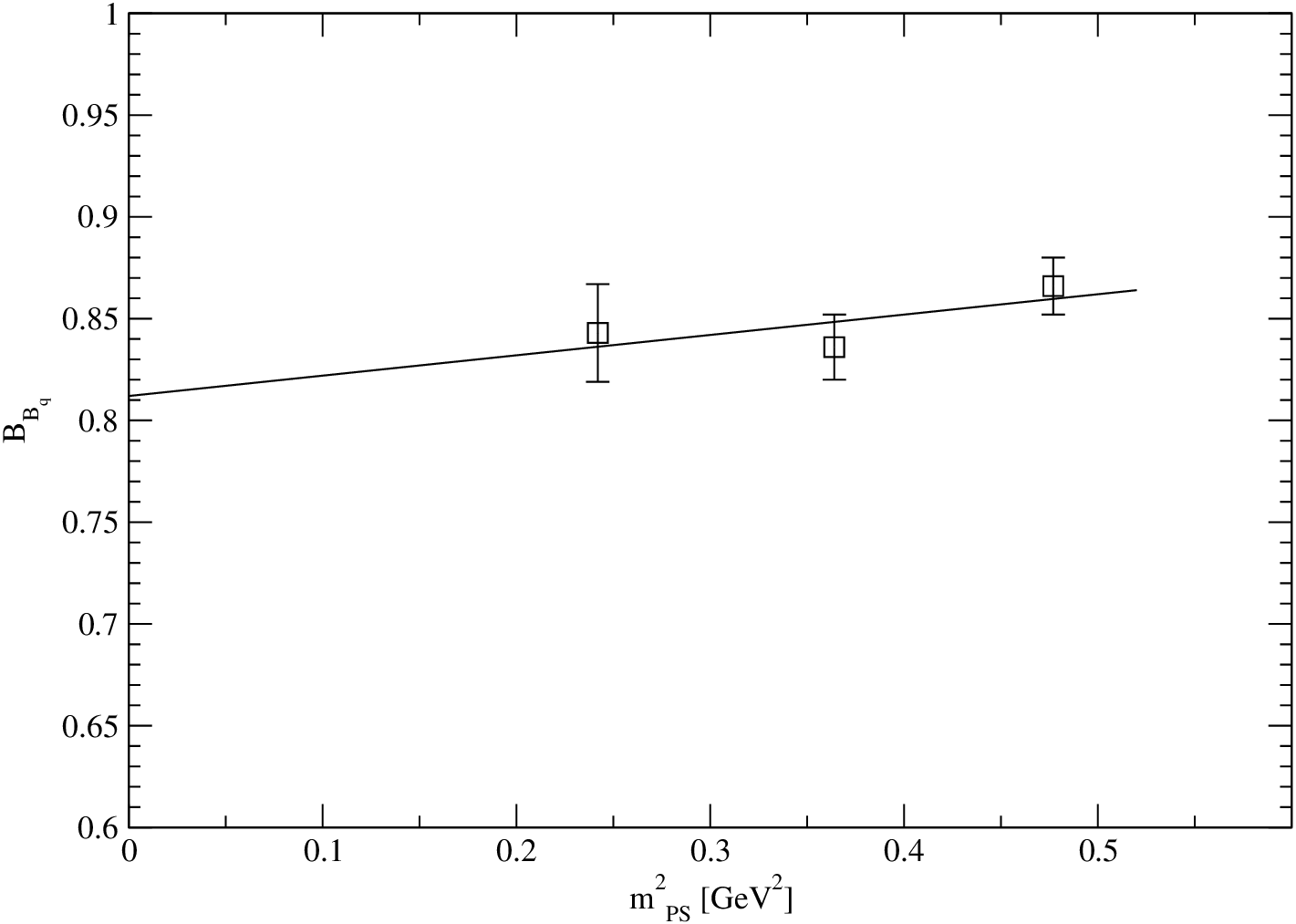}
\caption{\label{fig:b_b}Chiral extrapolation of $B_B$ obtained using Eq.~(\ref{eq:b_b_mix}). Error bars represent statistical errors.}
\end{figure}

\end{document}